\documentclass[11pt]{article}
\usepackage{moriond,epsfig}

\def\be{\begin{equation}}
\def\ee{\end{equation}}
\def\bea{\begin{eqnarray}}
\def\eea{\end{eqnarray}}

\def\Cas{\mathrm{Cas}}

\def\B{\mathrm{B}}

\def\bk{\mathbf{k}}

\def\dd{\mathrm{d}}

\def\TE{\mathrm{TE}}
\def\TM{\mathrm{TM}}

\def\P{\mathrm{P}}    

\def\max{\mathrm{max}}
\def\min{\mathrm{min}}
\def\perf{\mathrm{perf}}
\def\Gold{\mathrm{Gold}}
\def\PFA{\mathrm{PFA}}
\def\Tr{\mathrm{Tr}}
\def\calS{\mathcal{S}}
\def\calR{\mathcal{R}}
\def\calK{\mathcal{K}}
\def\calD{\mathcal{D}}
\def\calF{\mathcal{F}}

\begin{document}
\vspace*{4cm}
\title{CASIMIR AND SHORT-RANGE GRAVITY TESTS}

\author{A. LAMBRECHT and S. REYNAUD}

\address{Laboratoire Kastler Brossel, ENS, UPMC, CNRS, Jussieu, 75252 Paris, France\\
serge.reynaud@upmc.fr , www.lkb.ens.fr }

\maketitle\abstracts{ Comparison with theory of Casimir force
measurements are used to test the gravity force law at ranges from
0.1 to 10 micrometers. The interest of such tests depends crucially
on the theoretical evaluation of the Casimir force in realistic
experimental configurations. We present the scattering approach
which is nowadays the best tool for such an evaluation. We then
describe the current status of the comparisons between theory and
experiments. }

\section{Introduction}

The Casimir effect is an observable effect of quantum vacuum
fluctuations which deserves careful attention as a crucial
prediction of quantum field theory~\cite{Milonni94,LamoreauxResource99,%
Bordag01,Milton05,Barrera06,CasimirPhysics}.

Casimir physics also plays an important role in the tests of gravity
at sub-millimeter ranges~\cite{Fischbach98,Adelberger03}. Strong
constraints have been obtained in short range Cavendish-like
experiments~\cite{Kapner07,AdelbergerThisBook}. For scales of the
order of the micrometer, similar tests are performed by comparing
with theory the results of Casimir force
measurements~\cite{LambrechtPoincare,OnofrioNJP06}. At even shorter
scales, the same can be done with
atomic~\cite{Wolf07,Lepoutre09,PelleThisBook} or
nuclear~\cite{Nesvizhevsky08,NesvizhevskyThisBook} force
measurements. A recent overview of these short-range tests can be
found in~\cite{AntoniadisReview}.

In the following, we focus our attention on Casimir tests of the
gravity force law. They are performed at distances from 0.1 to 10
micrometers for which the Casimir force dominates the standard
gravity force. It follows that the hypothetical new force would be
seen as a difference between the experimental result
$F_\mathrm{exp}$ and theoretical prediction $F_\mathrm{th}$. This
implies that these two quantities have to be assessed independently
from each other. This situation should clearly forbid anyone to use
theory-experiment comparison to prove (or disprove) a specific
experimental result or theoretical model.

\section{The problem of vacuum energy}

Before entering this discussion, we want to emphasize that the
Casimir effect has a fascinating interface with the puzzles of
gravitational physics through the problem of vacuum
energy~\cite{GenetDark02,Jaekel08,JaekelThisBook}.

Nernst was the first physicist to notice as soon as in 1916 that
zero-point fluctuations of the electromagnetic field constituted a
challenge for gravitation theory~\cite{Nernst16,Browne95}. The very
existence of these fluctuations dismisses the classical idea of an
empty space. When the vacuum energy density is calculated by adding
the zero-point energies over all field modes, an infinite value is
obtained. When a high frequency cutoff is introduced, the sum is
finite but still much larger than the mean energy observed through
gravitational phenomena~\cite{Weinberg89,Adler95}.

This problem has led famous physicists to deny the reality of vacuum
fluctuations. In particular, Pauli crudely stated in his textbook on
Wave Mechanics~\cite{Pauli33} : \begin{quote} \emph{At this point it
should be noted that it is more consistent here, in contrast to the
material oscillator, not to introduce a zero-point energy of
$\frac12\hbar\omega$ per degree of freedom. For, on the one hand,
the latter would give rise to an infinitely large energy per unit
volume due to the infinite number of degrees of freedom, on the
other hand, it would be principally unobservable since nor can it be
emitted, absorbed or scattered and hence, cannot be contained within
walls and, as is evident from experience, neither does it produce
any gravitational field.} \end{quote}

A part of these statements is simply unescapable~: the mean value of
vacuum energy does not contribute to gravitation as an ordinary
energy. This is just a matter of evidence since the universe would
look very differently otherwise. But it is certainly no longer
possible to uphold today that vacuum fluctuations have no observable
effects. Certainly, vacuum fluctuations can \emph{be emitted,
absorbed, scattered...} as shown by their numerous effects in
atomic~\cite{CCT92} and subatomic~\cite{IZ85} physics. And the
Casimir effect~\cite{Casimir} is nothing but the physical effect
produced by vacuum fluctuations when they are \emph{contained within
walls}.

\section{The Casimir force in the ideal and real cases}

Casimir calculated the force between a pair of perfectly smooth,
flat and parallel plates in the limit of zero temperature and
perfect reflection. He found universal expressions for the force
$F_\Cas$ and energy $E_\Cas$
\begin{eqnarray}
F_\Cas=-\frac{\dd E_\Cas}{\dd L} \quad,\quad E_\Cas= - \frac{\hbar c
\pi^2 A}{720 L^3}
\end{eqnarray}
with $L$ the distance, $A$ the area, $c$ the speed of light and
$\hbar$ the Planck constant. This universality is explained by the
saturation of the optical response of perfect mirrors which reflect
100\% (no less, no more) of the incoming fields. In particular the
expressions $F_\Cas$ and $E_\Cas$ do not depend on the atomic
structure constants. Of course, this idealization is no longer
tenable for the real mirrors used in the experiments.

The effect of imperfect reflection is large in most experiments, and
a precise knowledge of its frequency dependence is essential for
obtaining a reliable theoretical prediction~\cite{LambrechtEPJ00}.
Meanwhile, experiments are performed at room temperature so that the
effect of thermal fluctuations has to be added to that of
vacuum~\cite{Mehra67,Schwinger78}. Then, precise experiments are
performed between a plane and a sphere whereas calculations are
often devoted to the geometry of two parallel planes. The estimation
of the force in the plane-sphere geometry involves the so-called
\textit{Proximity Force Approximation} (PFA)~\cite{Derjaguin68}
which amounts to averaging over the distribution of local
inter-plate distances the force calculated in the two-planes
geometry. But the PFA can only be valid when the radius $R$ is much
larger than the separation $L$ and even in this case its accuracy
has to be assessed.

\section{The calculation of the force in the scattering approach}

The best tool available for addressing these questions is the
scattering approach. This approach has been used for years for
evaluating the Casimir force between non perfectly reflecting
mirrors~\cite{Jaekel91,GenetPRA03}. It is today the best solution
for calculating the force in arbitrary geometries
\cite{LambrechtNJP06,Lambrecht11}.

The basic idea is that mirrors are described by their scattering
amplitudes. When studying first the geometry of two plane and
parallel mirrors aligned along the axis $x$ and $y$, these
amplitudes are specular reflection and transmission amplitudes ($r$
and $t$) which depend on frequency $\omega$, the transverse vector
$\bk \equiv \left( k_x,k_y\right)$ and the polarization $p=\TE,\TM$
(all these quantities being preserved by scattering). Two mirrors
form a Fabry-Perot cavity described by a global $S-$matrix which can
be evaluated from the elementary $S-$matrices associated with the
two mirrors. Thermal equilibrium is here assumed for the whole
system \emph{cavity + fields}. Care has to be taken to account for
the contribution of evanescent waves besides that of ordinary modes
freely propagating outside and inside the cavity. The properties of
the evanescent waves are described through an analytical
continuation of those of ordinary ones, using the well defined
analytic behavior of the scattering amplitudes. At the end of this
derivation, this analytic properties are also used to perform a Wick
rotation from real to imaginary frequencies.

The sum of the phaseshifts associated with all field modes leads to
the expression of the Casimir free energy $\calF$
\begin{eqnarray}
\label{CasimirFreeEnergy} &&\calF = \sum_\bk \sum_p k_\B T
\sum_m{}^\prime \ln d(i\xi_m,\bk,p) \quad,\quad d(i\xi,\bk,p) = 1 -
r(i\xi,\bk,p) e^{ -2\kappa L }
\\
&& r\equiv r_1 r_2 \quad,\quad \xi_m \equiv \frac{2\pi m k_\B
T}\hbar \quad,\quad \kappa \equiv \sqrt{\bk^2+\frac{\xi^2}{c^2}}
\nonumber
\end{eqnarray}
$\sum_\bk \equiv A \int\frac{\dd^2\bk}{4\pi^2}$ is the sum over
transverse wavevectors with $A$ the area of the plates, $\sum_p$ the
sum over polarizations and $\sum_m^\prime$ the Matsubara sum (sum
over positive integers $m$ with $m=0$ counted with a weight
$\frac12$); $r$ is the product of the reflection amplitudes of the
mirrors as seen by the intracavity field; $\xi$ and $\kappa$ are the
counterparts of frequency $\omega$ and longitudinal wavevector $k_z$
after the Wick rotation.

This expression reproduces the Casimir ideal formula in the limits
of perfect reflection $r \rightarrow 1$ and null temperature $T
\rightarrow 0$. But it is valid and regular at thermal equilibrium
at any temperature and for any optical model of mirrors obeying
causality and high frequency transparency properties. It has been
demonstrated with an increasing range of validity in
\cite{Jaekel91},~\cite{GenetPRA03} and~\cite{LambrechtNJP06}. The
expression is valid not only for lossless mirrors but also for lossy
ones. In the latter case, it accounts for the additional
fluctuations accompanying losses inside the mirrors.

It can thus be used for calculating the Casimir force between
arbitrary mirrors, as soon as the reflection amplitudes are
specified. These amplitudes are commonly deduced from models of
mirrors, the simplest of which is the well known Lifshitz model
\cite{Lifshitz56,DLP61} which corresponds to semi-infinite bulk
mirrors characterized by a local dielectric response function
$\varepsilon (\omega)$ and reflection amplitudes deduced from the
Fresnel law.

In the most general case, the optical response of the mirrors cannot
be described by a local dielectric response function. The expression
(\ref{CasimirFreeEnergy}) of the free energy is still valid in this
case with reflection amplitudes to be determined from microscopic
models of mirrors. Attempts in this direction can be found for
example in~\cite{Pitaevskii08,Dalvit08,Svetovoy08}.

\section{The case of metallic mirrors}

The most precise experiments have been performed with metallic
mirrors which are good reflectors only at frequencies smaller than
their plasma frequency $\omega_\P$. Their optical response is
described by a reduced dielectric function usually written at
imaginary frequencies $\omega=i\xi$ as
\begin{eqnarray}
\varepsilon \left[i\xi\right] = \hat{\varepsilon}\left[i\xi\right] +
\frac{\sigma \left[i\xi\right] }{\xi} \quad,\quad \sigma
\left[i\xi\right] = \frac{\omega_\P^2}{\xi+\gamma}
\end{eqnarray}
The function $\hat{\varepsilon} \left[i\xi\right] $ represents the
contribution of interband transitions and is regular at the limit
$\xi\to0$. Meanwhile $\sigma \left[i\xi\right]$ is the reduced
conductivity ($\sigma$ is measured as a frequency and the SI
conductivity is $\epsilon_0\sigma$) which describes the contribution
of the conduction electrons.

A simplified description corresponds to the lossless limit $\gamma
\to 0$ often called the plasma model. As $\gamma$ is much smaller
than $\omega_\P$ for a metal such as Gold, this simple model
captures the main effect of imperfect reflection. However it cannot
be considered as an accurate description since a much better fit of
tabulated optical data is obtained~\cite{LambrechtEPJ00} with a non
null value of $\gamma$. Furthermore, the Drude model $\gamma\neq0$
meets the important property of ordinary metals which have a finite
static conductivity
\begin{eqnarray}
\sigma_0 = \frac{\omega_\P^2}{\gamma}
\end{eqnarray}
This has to be contrasted to the lossless limit which corresponds to
an infinite value for $\sigma_0$.

When taking into account the imperfect reflection of the metallic
mirrors, one finds that the Casimir force is reduced with respect to
the ideal Casimir expression at all distances for a null
temperature~\cite{LambrechtEPJ00}. This reduction is conveniently
represented as a factor
\begin{eqnarray}
\eta_F = \frac{F}{F_\Cas} \quad,\quad F=-\frac{\partial
\calF}{\partial L}
\end{eqnarray}
where $F$ is the real force and $F_\Cas$ the ideal expression. For
the plasma model, there is only one length scale, tha plasma
wavelength $\lambda_P = 2\pi c/\omega_P $ in the problem (136nm for
Gold). The ideal Casimir formula is recovered ($\eta_F\to1$) at
large distances $L\gg\lambda_P$, as expected from the fact that
metallic mirrors tend to be perfect reflectors at low frequencies
$\omega \ll\omega_P$. At short distances in contrast, a significant
reduction of the force is obtained ($\eta_F\ll1$), which scales as
$L/\lambda_P$, as a consequence of the fact that metallic mirrors
are poor reflectors at high frequencies $\omega \gg\omega_P$. In
other words, there is a change in the power law for the variation of
the force with distance. This change can be understood as the result
of the Coulomb interaction of surface plasmons living at the two
matter-vacuum interfaces~\cite{Genet04,Intravaia05}.

As experiments are performed at room temperature, the effect of
thermal fluctuations has to be added to that of vacuum fields
\cite{GenetPRA00}. Significant thermal corrections appear at
distances $L$ larger than a critical distance determined by the
thermal wavelength $\lambda_T$ (a few micrometers at room
temperature). Bostr\"{o}m and Sernelius were the first to remark
that the small non zero value of $\gamma$ had a significant effect
on the force at non null temperatures~\cite{Bostrom00}. In
particular, there is a large difference at large distances between
the expectations calculated for $\gamma=0$ and $\gamma\neq0$, their
ratio reaching a factor 2 when $L\gg\lambda_T$. It is also worth
emphasizing that the contribution of thermal fluctuations to the
force is opposite to that of vacuum fluctuations for intermediate
ranges $L\sim\lambda_T$.

This situation has led to a blossoming of contradictory papers (see
references in~\cite{Reynaud03,BrevikNJP06,IngoldPRE09}). As we will
see below, the contradiction is also deeply connected to the
comparison between theory and experiments.

\section{The non-specular scattering formula}

We now present a more general scattering formula allowing one to
calculate the Casimir force between stationary objects with
arbitrary geometries. The main generalization with respect to the
already discussed cases is that the scattering matrix $\calS$ is now
a larger matrix accounting for non-specular reflection and mixing
different wavevectors and polarizations while preserving frequency.
Of course, the non-specular scattering formula is the generic one
while specular reflection can only be an idealization.

The Casimir free energy can be written as a generalization of
equation (\ref{CasimirFreeEnergy})
\begin{eqnarray}
\label{CasimirFreeEnergyNS}
&&\calF = k_\B T \sum_m{}^\prime \,\Tr \ln \calD (i\xi_m) \\
&&\calD = 1 - \calR_1 \exp^{ -\calK L } \calR_2 \exp^{ -\calK L }
\nonumber
\end{eqnarray}
The symbol $\Tr$ refers to a trace over the modes at a given
frequency. The matrix $\calD$ is the denominator containing all the
resonance properties of the cavity formed by the two objects 1 and 2
here written for imaginary frequencies. It is expressed in terms of
the matrices $\calR_1$ and $\calR_2$ which represent reflection on
the two objects 1 and 2 and of propagation factors $\exp^{-\calK
L}$. Note that the matrices $\calD$, $\calR_1$ and $\calR_2$, which
were diagonal on the basis of plane waves when they described
specular scattering, are no longer diagonal in the general case of
non specular scattering. The propagation factors remain diagonal in
this basis with their diagonal values written as in
(\ref{CasimirFreeEnergy}). Clearly the expression
(\ref{CasimirFreeEnergyNS}) does not depend on the choice of a
specific basis. But it may be written in specific basis fitting the
geometry under study.

The multiple scattering formalism has been used in the past years by
different groups using different notations (see as examples
\cite{Emig08,Kenneth08,Milton08}) and numerous applications have
been considered. In particular, the case of corrugated plates or
gratings has been extensively studied
\cite{Rodrigues06,Rodrigues07,RodriguesEPL06,Lambrecht08} and it has
given rise to interesting comparisons with experiments
\cite{Chan08,Chiu09,Bao10}. Note also that calculations have been
devoted to the study of atoms in the vicinity of corrugated
plates~\cite{DalvitPRL08,MessinaPRA09,ContrerasPRA10}.

\section{The plane-sphere geometry beyond PFA}

Recently, it has also become possible to use the general scattering
formula to obtain explicit evaluations of the Casimir force in the
plane-sphere geometry.  Such calculations have first been performed
for perfectly reflecting mirrors~\cite{Maia08}. They have then been
done for the more realistic case of metallic mirrors described by a
plasma model dielectric function~\cite{CanaguierPRL09}. Even more
recently, calculations were made which treat simultaneously
plane-sphere geometry and non zero temperature, with dissipation
taken into account~\cite{CanaguierPRL10}.

In these calculations, the reflection matrices are written in terms
of Fresnel amplitudes for plane waves on the plane mirror and of Mie
amplitudes for spherical waves on the spherical mirror. The
scattering formula is then obtained by writing also transformation
formulas from the plane waves basis to the spherical waves basis and
conversely. The energy takes the form of an exact multipolar formula
labeled by a multipolar index $\ell$. When doing the numerics, the
expansion is truncated at some maximum value $\ell_\max$, which
degrades the accuracy of the resulting estimation for very large
spheres $x\equiv L/R<x_\min$ with $x_\min$ proportional to
$\ell_\max^{-1}$.

The results of these calculations may be compared to the
experimental study of PFA in the plane-sphere
geometry~\cite{Krause07}. In this experiment, the force gradient is
measured for various radii of the sphere and the results are used to
obtain a constraint $\vert\beta_G\vert<0.4$ on the slope at origin
$\beta_G$ of the function $\rho_G(x)$
\begin{eqnarray}
\rho_G=\frac{G}{G^\PFA}=1+\beta_G x+O(x^2) \quad,\quad x\equiv \frac
LR
\end{eqnarray}
The slope obtained by interpolating at low values of $x$ our
theoretical evaluation of $\rho_G$ reveals a striking difference
between the cases of perfect and plasma mirrors. The slope
$\beta_G^\perf$ obtained for perfect mirrors is larger than that
$\beta_G^\Gold$ obtained for gold mirrors by a factor larger than 2
\begin{eqnarray}
\beta_G^\perf \sim-0.48 \quad,\quad \beta_G^\Gold \sim-0.21
\end{eqnarray}
As a result, $\beta_G^\Gold$ is compatible with the experimental
bound whereas $\beta_G^\perf$ is not~\cite{CanaguierPRL09}.

The effect of temperature is also correlated with the plane-sphere
geometry. The first calculations accounting simultaneously for
plane-sphere geometry, temperature and dissipation have been
published very recently~\cite{CanaguierPRL10} and they show several
striking features. The factor of 2 between the long distance forces
in Drude and plasma models is reduced to a factor below 3/2 in the
plane-sphere geometry. Then, PFA underestimates the Casimir force
within the Drude model at short distances, while it overestimates it
at all distances for the perfect reflector and plasma model. If the
latter feature were conserved for the experimental parameter region
$R/L$ $(>10^2)$, the actual values of the Casimir force calculated
within plasma and Drude model could turn out to be closer than what
PFA suggests. This would affect the discussion of the next section,
which is still based on calculations using PFA.

\section{Discussion of experiments}

We end up this review by discussing the status of comparisons
between Casimir experiments and theory. We emphasize that, after
years of improvement in experiments and theory, we have to face a
lasting discrepancies in their comparison.

On one side, the Purdue and Riverside experiments
\cite{DeccaAP05,DeccaPRD07,KlimchitskayaRMP09} appear to favor
predictions obtained with $\gamma=0$ rather than those corresponding
to the expected $\gamma\neq0$ (see Fig.1 in~\cite{DeccaPRD07}). This
result stands in contradiction to the  fact that Gold has a finite
conductivity. Note that these experiments are done at distances
smaller than 0.75$\mu$m where the thermal contribution is small, so
that accuracy is a critical issue here.

On the other side, a new experiment at Yale \cite{SushkovNatPh11}
has been able to measure the force at larger distances
(0.7$\mu$m-7$\mu$m) where the thermal contribution is larger and the
difference between the predictions at $\gamma=0$ and $\gamma\neq0$
significant. The results favor the expected Drude model
($\gamma\neq0$), but only after subtraction of a large contribution
of the patch effect.

It is worth emphasizing that the results of the new experiment see a
significant thermal contribution and fit the expected model. Of
course, they have to be confirmed by further studies
\cite{MiltonNatPh11}. In particular, the electrostatic patch effect
remains a source of concern in Casimir experiments
\cite{SpeakePRL03,KimPRA10}. It is not measured independently in any
of the experiments discussed above. This means that the Casimir
effect, which is now verified in several experiments, is however not
tested at the 1\% level, as has been sometimes claimed. This also
entails that the tests of gravity at the micrometer range have still
room available for improvement.

\section*{Acknowledgments}
The authors thank A. Canaguier-Durand, A. G\'erardin, R. Gu\'erout,
J. Lussange, R.O. Behunin, I. Cavero-Pelaez, D. Dalvit, C. Genet,
G.L. Ingold, F. Intravaia, M.-T. Jaekel, P.A. Maia Neto, V.V.
Nesvizhevsky for contributions to the work reviewed in this paper,
and the ESF Research Networking Programme CASIMIR
(www.casimirnetwork. com) for providing excellent opportunities for
discussions on the Casimir effect and related topics.

\end{document}